\documentclass[conference]{IEEEtran}
\usepackage[a4 paper, top=0.75in, bottom=0.75in, left=0.75in, right=0.75in]{geometry}
\usepackage{graphicx}
\usepackage{amsfonts}
\usepackage{amsthm}
\usepackage{color}
\usepackage{amsmath}
\usepackage{subfigure}
\usepackage{algorithm}
\usepackage{epstopdf}
\usepackage{array}
\usepackage{amssymb}
\usepackage{enumitem}
\usepackage{epstopdf}
\usepackage{stfloats}
\usepackage{dsfont}  
\usepackage{mdwmath}
\usepackage{mdwtab}
\usepackage{mathtools}
\usepackage{multirow}

\ifCLASSINFOpdf
\else
\fi
\usepackage{cite}

\begin{document}
\hyphenation{op-tical net-works semi-conduc-tor}
\renewcommand{\thefootnote}{}
\title{On the Performance of PDCCH in LTE and 5G New Radio} 

\author{\IEEEauthorblockN{Hongzhi~Chen\IEEEauthorrefmark{1}, De~Mi\IEEEauthorrefmark{1}, Manuel~Fuentes\IEEEauthorrefmark{4}, 
		Eduardo~Garro\IEEEauthorrefmark{4}, Jose~Luis Carcel\IEEEauthorrefmark{2}, \\ Belkacem~Mouhouche\IEEEauthorrefmark{2}, Pei~Xiao\IEEEauthorrefmark{1} and Rahim~Tafazolli\IEEEauthorrefmark{1}}
	\IEEEauthorblockA{\IEEEauthorrefmark{1}Institute for Communication Systems, University of Surrey, United Kingdom}
	
	\IEEEauthorblockA{\IEEEauthorrefmark{4}Institute of Telecommunications and Multimedia Applications, Universitat Politecnica de Valencia, Spain}
	\IEEEauthorblockA{\IEEEauthorrefmark{2}Samsung Electronics R{\&D} UK, United Kingdom}
	Email:\{hongzhi.chen, d.mi, p.xiao, r.tafazolli\}@surrey.ac.uk, \{mafuemue, edgarcre\}@iteam.upv.es, \\ \{jose.cervera, b.mouhouche\}@samsung.com}


%


\maketitle
\begin{abstract}
5G New Radio (NR) Release 15 has been specified in June 2018. It introduces numerous changes and potential improvements for physical layer data transmissions, although only point-to-point (PTP) communications are considered. In order to use physical data channels such as the Physical Downlink Shared Channel (PDSCH), it is essential to guarantee a successful transmission of control information via the Physical Downlink Control Channel (PDCCH). Taking into account these two aspects, in this paper, we first analyze the PDCCH processing chain in NR PTP as well as in the state-of-the-art Long Term Evolution (LTE) point-to-multipoint (PTM) solution, i.e., evolved Multimedia Broadcast Multicast Service (eMBMS). Then, via link level simulations, we compare the performance of the two technologies, observing the Bit/Block Error Rate (BER/BLER) for various scenarios. The objective is to identify the performance gap brought by physical layer changes in NR PDCCH as well as provide insightful guidelines on the control channel configuration towards NR PTM scenarios. 
\end{abstract}


%
\IEEEpeerreviewmaketitle

{\scriptsize \footnote{We would like to acknowledge the support of the University of Surrey 5GIC (www.surrey.ac.uk/5gic) members for this work. This work was also supported in part by the European Commission under the 5GPPP project 5G-Xcast (H2020-ICT-2016-2 call, grant number 761498). The views expressed in this contribution are those of the authors and do not necessarily represent the project.}}


\section{Introduction}
Driven by an unprecedented growth of the traffic demand and multimedia content consumption, the 3rd Generation Partnership Project (3GPP) specified in Long Term Evolution (LTE) from Release (Rel-) 9 the use of point-to-multipoint (PTM) for broadcast and multicast via evolved Multimedia Broadcast Multicast Service (eMBMS). Currently, the state-of-the-art specification for eMBMS is LTE-Advanced Pro Rel-14. This PTM service allows the content provider to efficiently deliver services to a large group of users who are interested in the same media content, with a fixed amount of radio resources. The performance of the latest eMBMS solution has been analyzed in our prior work \cite{SAM}\cite{TBC}, focusing on data transmissions. Most recently, the Rel-15 of fifth-generation (5G) New Radio (NR) specification has been released, which brings a lot of physical layer changes. It is then crucial to investigate how those changes can affect the performance of PTM transmissions in 5G.

Authors in \cite{NRDATA} have studied the traffic channel performance based on NR specification, focusing on the multimedia broadcast and multicast services. However, the overall system performance should also consider the control channel. For example, the probability of radio link failure will increase if the Block Error Rate (BLER) of the Downlink Control Information (DCI) exceeds a certain target threshold. Robustness is the key design principle for control channels and very robust Forward Error Correction (FEC) codes and small modulation orders ensure that the control information is received correctly. In the literature, the performance of LTE control channels with fixed receivers and perfect channel estimators has been studied in \cite{PCFICHPDCCHLTE}, and a potential power-based optimization for the control information transmission based on LTE has been discussed in \cite{powerdci}.

In this work, we present a comprehensive technical overview of both LTE eMBMS and NR point-to-point (PTP) systems in order to provide insightful guidelines on the control channel configuration towards NR PTM scenarios. Note that the first release of 5G, i.e. Rel-15 only specifies the use of unicast, but it can be used as a basis for evaluating a possible NR PTM solution. After describing the DCI processing chain for both systems in detail, a performance analysis is provided via link-level simulations. It is based on the evaluation methodology defined by the International Communications Union - Recommendation (ITU-R) for the International Mobile Telecommunication 2020 (IMT-2020) evaluation process \cite{ITU_R_Guidelines}. It follows the physical layer chain defined by 3GPP in \cite{TR36212} and \cite{TR38211}. The obtained results can be used as a comparison of the current LTE PTM solution and NR PTP including the physical layer changes specified in Rel-15. This analysis can be also extrapolated to the evaluation of an end-to-end system performance including the transmission on data channels e.g. Physical layer Downlink Shared Channel (PDSCH), or in the case of proposing a suitable control channel configuration for NR PTM scenarios.

This paper is structured as follows. First, Section II describes the DCI formats for both LTE-eMBMS and NR-PTP technologies. Their individual transmit block diagrams for PDCCH are given in Section III. Section IV presents and discusses their corresponding frame structure. PDCCH link-level simulation results in various scenarios are included in Section V. Finally, Section VI concludes the key findings and discusses the potential improvements towards the development of a technical solution for NR PTM in the future.

\section{Downlink Control Information Generation}
Different types of information can be transmitted inside the physical control channels, as Table \ref{table:1} specifies. Control information for one or multiple User Equipments (UEs) is located in a single DCI message and is transmitted through the PDCCH. Different DCI formats are defined for specific purposes. From the eMBMS content transmission point of view, there is no support for the Hybrid Automatic Repeat Request (HARQ) operation, and therefore the transmission of CFI symbols works similarly as the DCI information \cite{PCFICHPDCCHLTE} \cite{CFI} does. Therefore, in this paper, we focus on the signal processing and the corresponding performance of DCIs included in both LTE-eMBMS and NR-PTP configurations. 

\begin{table}[t]
	\centering
	\caption{Control information and corresponding channels}
	\renewcommand{\arraystretch}{1.2}
	\begin{tabular}{||c|c||} 
		\hline
		Control Information & Physical Control Channel \\
		\hline\hline
		Downlink Control Information (DCI)& PDCCH  \\ 
		Control Format Indicator (CFI) & PCFICH \\
		Hybrid-ARQ Indicator (HI) & PHICH\\
		\hline
	\end{tabular}
	\label{table:1}
\end{table}

\subsection{Long Term Evolution}
In LTE, different DCI formats contain diverse information, including Resource Block (RB) assignment, Transmit Power Commands (TPC), HARQ, precoding information, etc. Two significant factors that determine the format used for a specific situation are \cite{MRNTI}:

\begin{itemize}
	\item Radio Network Temporary Identifier (RNTI) type.
	\item Different transmission modes.
\end{itemize}

An example is given in Table \ref{table:2} \cite{TR36212}, related to the MBMS specific RNTI indication, called M-RNTI \cite{MRNTI}. The DCI format 1C (common search space) with M-RNTI is used for notification and includes an 8-bit bitmap to indicate the one or more Single Frequency Network (SFN) areas in which the Multicast Control Channel (MCCH) is used. In Table \ref{table:2}, the occupied channel bandwidth is also included.

\begin{table}[t]
	\centering
	\caption{DCI Format 1C for M-RNTI in LTE}
		\renewcommand{\arraystretch}{1.2}
	\begin{tabular}{||c|c||} 
		\hline
		Field Names& Occupied Bits \\
		\hline\hline
		MCCH Change Notification & 8 bits  \\ 
		\multirow{6}{*}{Reserved}&N/A~ (1.4MHz)\\&2bits~(3MHz)\\&4bits~(5MHz)\\&5bits~(10MHz)\\&6bits~(15MHz)\\&7bits~(20MHz)\\
		\hline
	\end{tabular}
	\label{table:2}
\end{table}
 
\subsection{New Radio}
In 5G NR, even in the current latest version of the 3GPP document TS 38.212 \cite{TS38212}, there is no specific DCIs defined for eMBMS transmissions. The current available DCI formats specified for PDSCH scheduling are shown in Table  \ref{table:3}.

\begin{table}[t]
	\centering
	\caption{DCI Formats in NR for PDSCH scheduling}
		\renewcommand{\arraystretch}{1.2}
	\begin{tabular}{||c|c||} 
		\hline
		Format & Usage \\
		\hline\hline
		Format 1$\rule{0.1cm}{0.15mm}$0 & used for the scheduling of PDSCH in one DL cell \\ 
		Format 1$\rule{0.1cm}{0.15mm}$1 & used for the scheduling of PDSCH in one cell\\
		\hline
	\end{tabular}
	\label{table:3}
\end{table}

\noindent where Format 1$\rule{0.1cm}{0.15mm}$0 is more suitable for a multicast/broadcast case as it is dedicated for DL cells. The specific fields that are included in this format are given in Table \ref{table:4}. In this table, VRB and PRB represent Virtual Resource Block and Physical Resource Block, respectively. With this format, the Cyclic Redundancy Check (CRC) is scrambled by C-RNTI since it is only defined for unicast in the current specification. Moreover, some of the fields do not have strong relations with eMBMS, such as those fields related to uplink control channels and for HARQ retransmissions. Besides, no format supports CRC scrambled by M-RNTI (MBMS-RNTI) or G-RNTI (group-RNTI).

\begin{table}[t]
	\centering
	\caption{DCI Format 1\_0 for C-RNTI in NR}
		\renewcommand{\arraystretch}{1.2}
	\begin{tabular}{||c|c||} 
		\hline
		Field Names& Occupied Bits \\
		\hline\hline
		Identifier for DCI formats & 1 bits  \\ 
	    Frequency domain resource assignments & Variable \\
	    Time domain resource assignments & X bits \\
	    VRB-to-PRB mapping & 1 bits  \\
	    Modulation and Coding scheme & 5 bits  \\
	    New data indicator & 1 bits  \\
	    Redundancy version & 2 bits  \\
	    HARQ process number & 4 bits  \\
	    Downlink assignment index & 2 bits  \\
	    TPC command assignment for scheduled PUCCH & 2 bits  \\
	    PUCCH resource indicator & 3 bits  \\
	    PDSCH-to-HARQ feedback timing indicator & 3 bits  \\
		\hline
	\end{tabular}
	\label{table:4}
\end{table}

The chosen corresponding DCI bits are then sent to the PDCCH channel processing chain, which is explained in next section.

\section{Physical Downlink Control Channels}
\subsection{Long Term Evolution}
DCI bits for different users are individually sent through the Bit-Interleaved Coding and Modulation (BICM) processing chain. DCI bits are encoded with a combination of forwarding error correction (FEC), scrambler and modulator. More specifically, a CRC sequence with 16 bits is first attached to the DCI information. Then, in the channel coding block, conventional tail-biting encoding (with code rate $R=1/3$) is employed. Next, rate matching is performed such that the bits inside each coding block are interleaved, circular buffered and punctured/repeated to provide a specific code rate (CR). The CR is determined by the Aggregation Level (AL). After rate matching, PDCCH bits for different users are multiplexed and scrambled before sending to the modulator. It is noticeable that the available modulation scheme for LTE PDCCH is Quadrature Phase Shift Keying (QPSK) only, as the transmission reliability of DCI bits is much more important than the transmission rate. After that, symbols are allocated in the available Resource Elements (RE) in the corresponding subframe, and finally, before transmission, CP-OFDM (Cyclic Prefix-Orthogonal Frequency Division Multiplexing) is performed.

In LTE, the PDCCH is categorized into common and UE-specific PDCCHs. Each type supports a particular set of searching spaces. Each searching space consists of a group of consecutive Control Channel Elements (CCE), which can be allocated in a specific PDCCH, called a PDCCH candidate. The most important LTE resource allocation units are then:

\begin{itemize}
	\item Resource Element (RE);
	\item Resource Element Group (REG);
	\item Control Channel Element (CCE);
	\item Aggregation Level (AL).
\end{itemize}	

In LTE, 1 CCE is formed by 9 REGs and 1 REG is formed in turn by 4 REs. Moreover, AL denotes the number of CCEs that carries a single PDCCH. If we assume that the AL is 1, then the total number of available REs for the whole control region is given by:

\begin{equation}\label{NRLTE}
\begin{aligned}
\text{RE}_\text{tot,LTE} &= \text{AL}*(\text{REG/CCE})*(\text{RE/REG})\\
& = 1*9*4 = 36 \text{REs},
\end{aligned}
\end{equation} 

where $\text{REG/CCE}$ represents the relationship between REG and CCE, and $\text{RE/REG}$ is the relation between RE and REG.

\subsection{New Radio}

\begin{figure*}[t]
	\centering
	\includegraphics[width=1.0\textwidth]{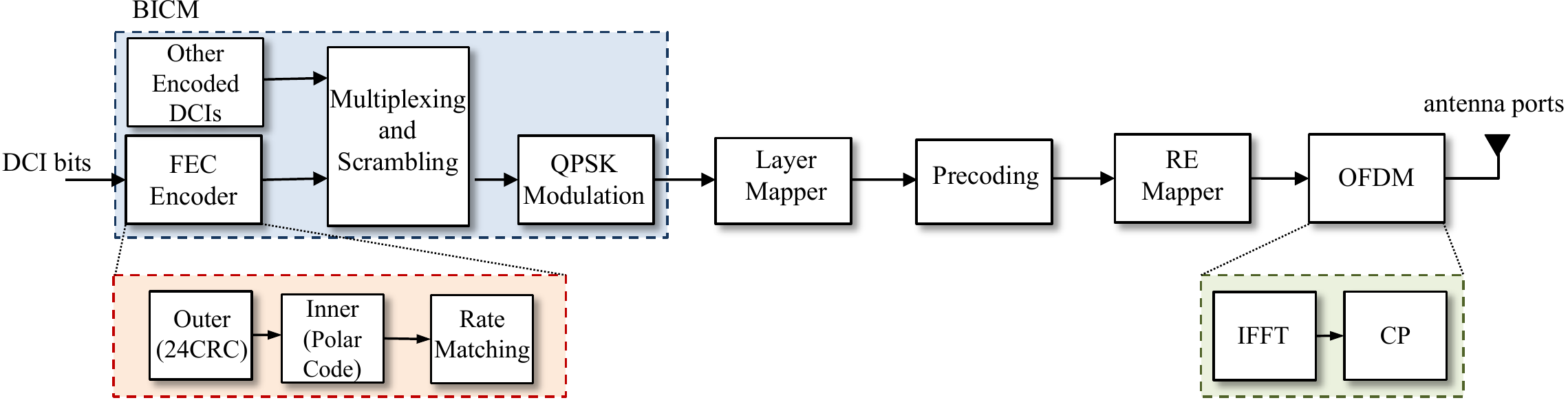}
	\caption{NR physical layer point-to-point PDCCH transmit block diagram.}
	\label{blockdiagram}
\end{figure*}

In NR, the PDCCH processing chain is different. The transmit block diagram of the NR PDCCH processing chain is shown in Fig.\ref{blockdiagram}. In this case, the generator polynomial $g_\text{CRC24C}(D)$ is used for the Cyclic Redundancy Check Attachment. Polar codes are used for channel coding. The specific details of polar encoding in NR can be found in \cite{TR38211}. The length of the polar encoded bits is $N=2^n$, where the $n$ is a positive integer between 5 and 9, both included. Therefore, the maximum length of the encoded bits is $2^9=512$. Regarding the rate matching, it is still operated every coded block and consists of sub-block interleaving, bit collection, and bit interleaving. However, according to the 3GPP specifications, the bit interleaving option is set to 0. The multiplexing and scrambling operations are the same than in 4G LTE. The same occurs with the modulation scheme, i.e., only QPSK is used.

Similarly to LTE, each PDCCH is still flexibly mapped to CCEs, but the relationship between REG and CCE changes in NR. In this case, 1 CCE is now formed by 6 REGs and 1 REG consists of 1 RB, which is equivalent to 12 REs in the frequency domain and 1 OFDM symbol in time domain. In this case, if we still assume the AL to be 1, the total number of available REs for the whole control region in NR is given by:

\begin{equation}\label{NRRE}
\begin{aligned}
\text{RE}_\text{tot,NR} &= \text{AL}*(\text{REG/CCE})*(\text{RE/REG})\\
& = 1*6*12 = 72 \text{REs},
\end{aligned}
\end{equation} 

\begin{figure}[t]
	\centering
	\includegraphics[width=0.49\textwidth]{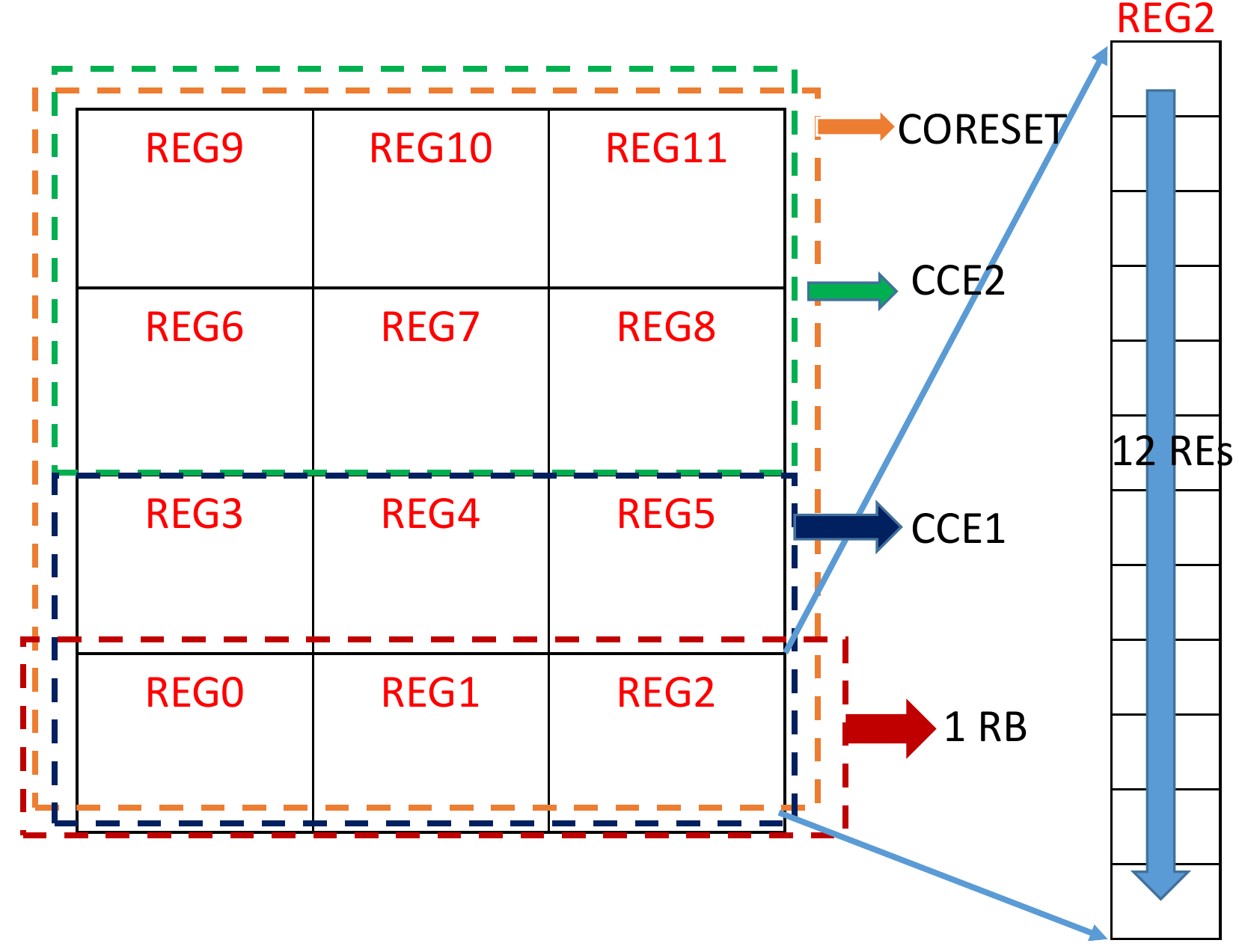}
	\caption{CORESET structure when considering 2 CCEs.}
	\label{Fig_CORESET}
\end{figure}

Note that some new units are defined in NR, including a REG bundle, which consists of multiple REGs, and the concept of a Control Resource Set (CORESET). A CORESET is made up of multiples RBs in the frequency domain and 1, 2 or 3 OFDM symbols in the time domain. CORESETs are equivalent to the control region in LTE subframes. A UE can be configured with multiple CORESETs, and each CORESET is associated to a single CCE-to-REG mapping only \cite{TR38211}. Fig.~\ref{Fig_CORESET} gives an example of the mapping from RE and REG to CCE with aggregation level 1. In this example, 3 OFDM symbols are used in the time domain, and there are 2 CCEs in the CORSET.

\section{Resource Block Structure and Code Rate Calculation}\label{sec:FS}

\begin{figure}[t]
	\centering
	\includegraphics[width=0.49\textwidth]{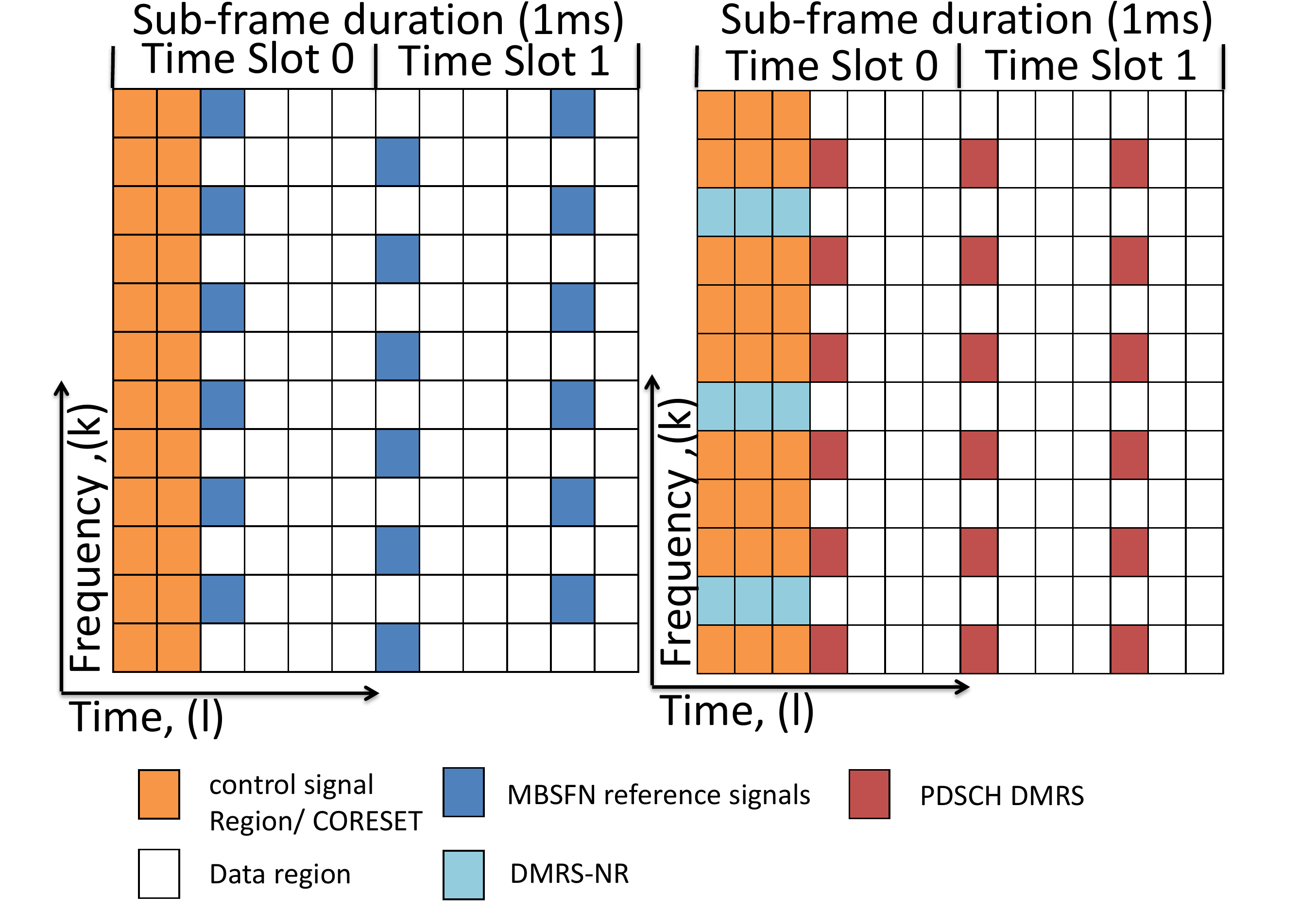}
	\caption{Resource Block structure in LTE-MBSFN (left) and NR (right), with 15 kHz carrier spacing.}
	\label{Fig_Framestructure}
\end{figure} 

Fig. \ref{Fig_Framestructure} shows the two possible frame structures for a single RB with LTE-eMBMS and NR \cite{TR38211}. As Fig. \ref{Fig_Framestructure} depicts, the number of subcarriers per RB are 12 in both cases. LTE-eMBMS permits the transmission of a specific modes called MBMS over Single Frequency Networks (MBSFN). This mode permits the use of 3 different configurations, corresponding to subcarrier spacings of 15 kHz, 7.5 kHz and 1.25 kHz. It also uses an exteded CP. On the other hand, NR permits the use of different numerologies related to specific carrier spacing. In particular, it is possible to use 15, 30, 60, 120 or 240 KHz. In this paper we focus on the subcarrier spacing of 15 kHz for both technologies, as the numerology that both have in common. The number of OFDM symbols per subcarrier with MBSFN and NR is 12 and 14 respectively. This is use to the use of extended and normal CP. It is worth mentioning that there is no reference signal in the control region for LTE while 9 Demodulation Reference Signal (DMRS) subcarriers are placed in the control region for NR.

Next, we provide an example of the calculation of the effective code rate for PDCCH. Since our focus in this paper is the performance of the control channel from a multicast/broadcast point of view, we also assume that there are no bits used for PHICH, and 2 and 3 bits are employed for PCFICH for LTE and NR respectively. The DCI bits for both LTE and NR are set to 12. In the case of LTE, 12 bits DCI can be seen as a Format 1C with 5 MHz of channel bandwidth. For NR, since there is no specific format suitable for multicast/broadcast and considering the fact that the smallest number of DCI bits is 12 \cite{TS38212}, we can keep the same size for DCI here. If we take into account these assumptions is (\ref{NRLTE}) and (\ref{NRRE}), with QPSK modulation, the total available bits will be 72 and 144 respectively. It means that 2 RBs are needed in both LTE and NR to allocate all necessary PDCCH symbols. Therefore, the effective code rate for LTE and NR under these assumptions is:

\begin{equation}
\text{CR}_\text{LTE} = \frac{12}{72}\approx 0.167
\end{equation}
 \begin{equation}
 \text{CR}_\text{NR} = \frac{12}{144-\text{bits}_\text{DMRS}}=\frac{12}{144-9*2} \approx 0.095
 \end{equation} 
 
\vspace{0.5cm}
 
\section{Link-Level Simulation Evaluation}\label{sec:simulation}
Link-level results, provided as the carrier-to-noise ratio (CNR) required to provide a specific BER and BLER value, are presented in this section for both LTE-eMBMS and NR. Different channel models, i.e. AWGN, TDL-A and TDL-C have been evaluated in order to better assess the impact of the  adopted configurations. Power Delay Profiles (PDP) for TDL-A and TDL-C channel are available in \cite{TR38901} and the simulation parameters are listed in Table \ref{table:paramater}.

\begin{table}[t]
	\centering
	\caption{Simulation Parameters}
	\begin{tabular}{||c| c||} 
		\hline
		Samulation Parameters & Values \\
		\hline\hline
		Carrier Frequency & 700 MHz \\ 
		System bandwidth & 5 MHz \\ 
		FFT size & 512 \\ 
		\multirow{2}{*}{CP types}&Extended for LTE-eMBMS\\& Normal for NR-PTP\\
	    DCI length & 12 bits \\
	    Aggregation Level & 1,2,4,8 \\
	    Subcarrier spacing & 15 kHz\\
	    Channel models & AWGN, TDL-A, TDL-C\\
	    \multirow{2}{*}{Channel estimation}&Perfect or 2-dimensional pilot based \\& estimation with linear interpolation\\
	    Equalizer & Minimum mean square error \\
		\hline
	\end{tabular}
	\label{table:paramater}
\end{table}

It is important to note that aggregation level 16 is not considered in this work, since it requires $16 \cdot 6 \cdot 12/(12\cdot3)=32$ REs and in a 5 MHz channel the maximum available REs are 25. But in theory, as the code rate for AL16 is halved compared to AL8 therefore, the expected required CNR value for BLER $10^-3$ with AL16 should be 3dB less than AL8. We use the AWGN channel to compare the performance of LTE-eMBMS and NR-PTP, but for TDL-A and TDL-C channels only NR has been considered.

\subsection{Ideal Channel Estimation}
In this section, we assume a standstill receiver with perfect channel estimation. Moreover, at the receiver side, the rate recovery process includes additively combining any repetitions to distinguish the performance difference of higher aggregation levels.

\subsubsection{Additive White Gaussian Noise Channel}
\begin{figure}[t]
	\centering
	\includegraphics[width=0.49\textwidth]{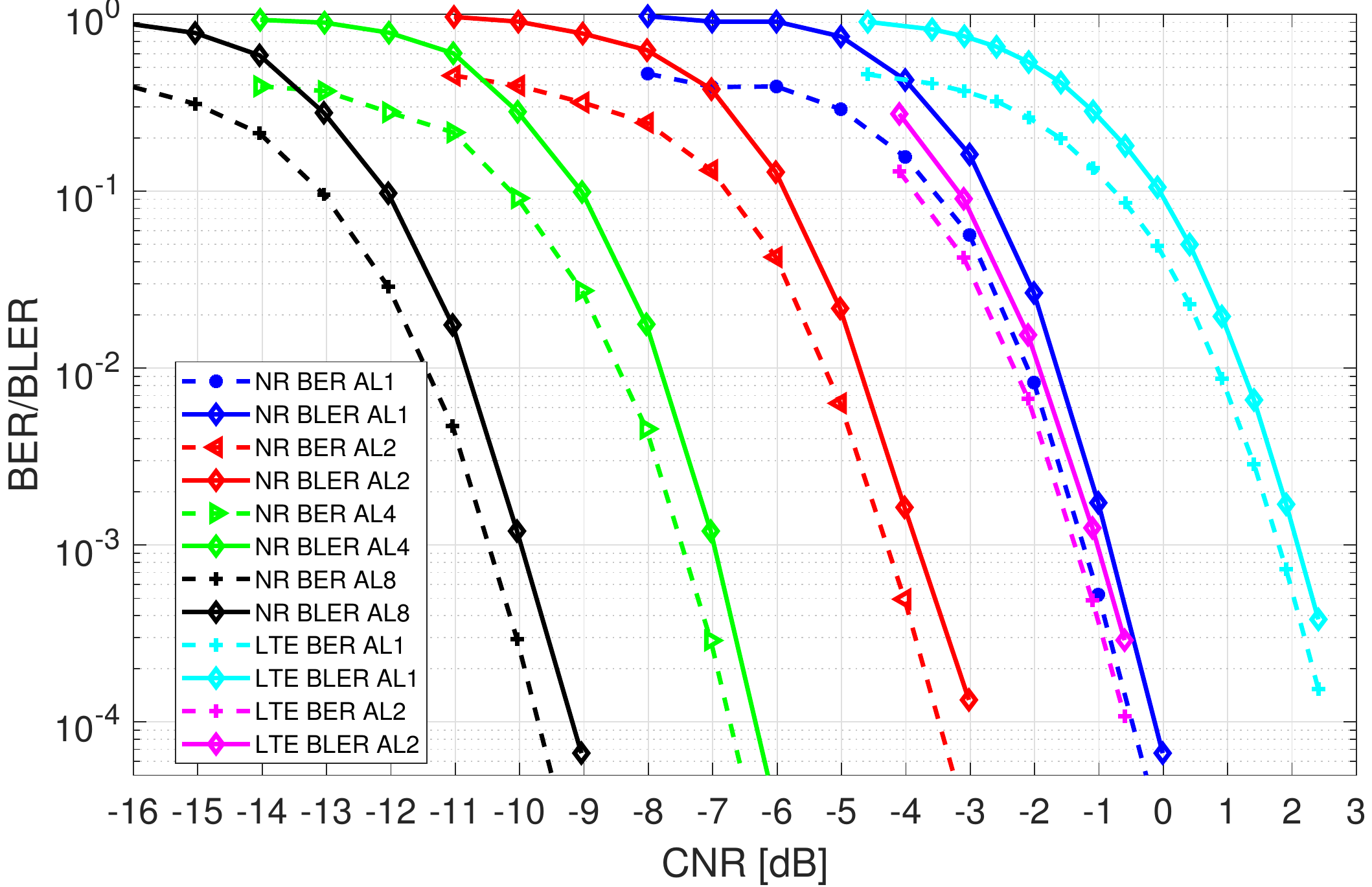}
	\caption{BER/BLER vs. CNR (dB) of LTE eMBMS and 5G NR for AWGN, ideal channel estimation.}
	\label{AWGNall}
\end{figure} 
\begin{figure}[t]
	\centering
	\includegraphics[width=0.49\textwidth]{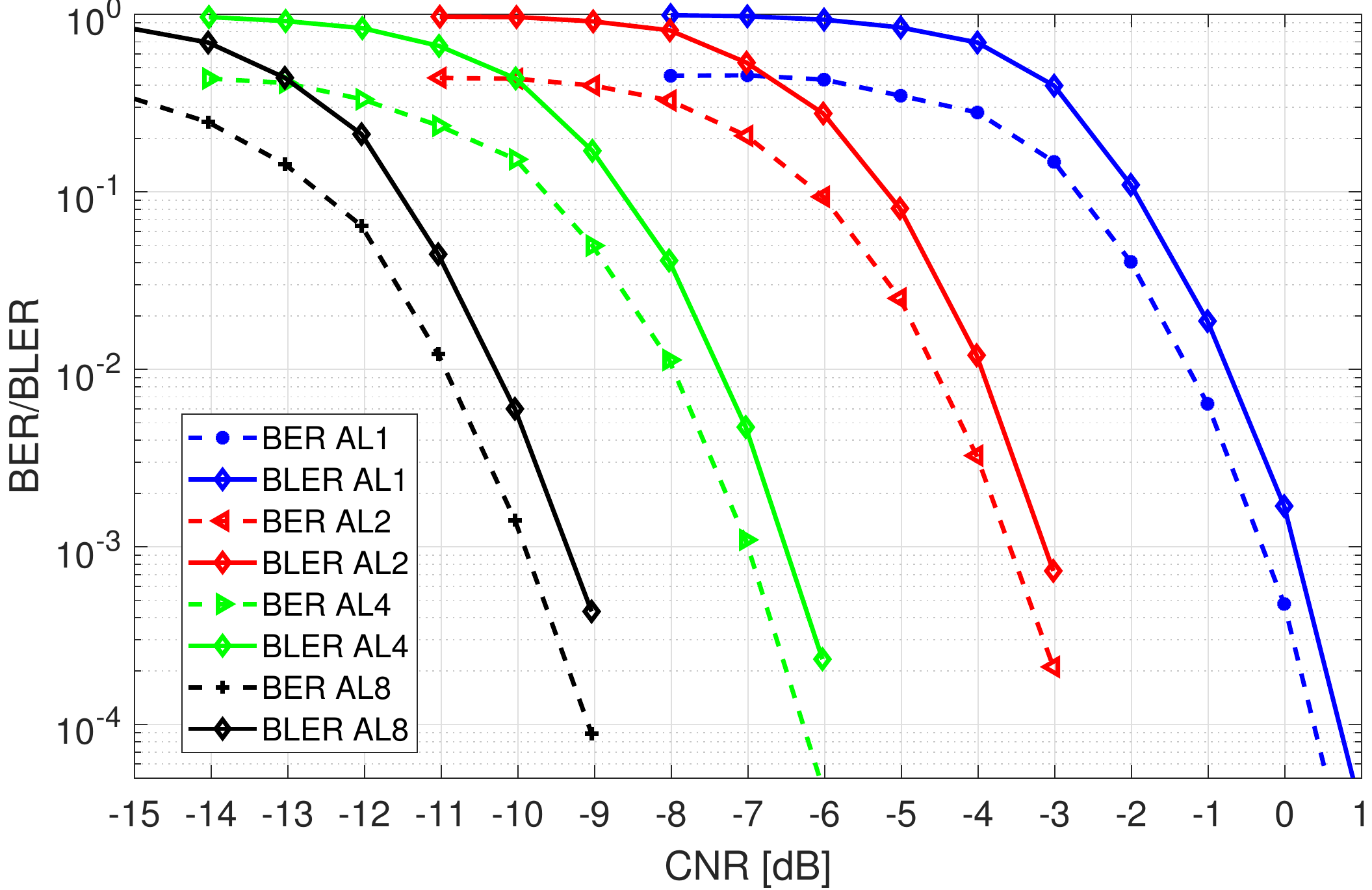}
	\caption{BER/BLER vs. CNR (dB) for TDL-A, ideal channel estimation.}
	\label{CDLA}
\end{figure} 
\begin{figure}[t]
	\centering
	\includegraphics[width=0.49\textwidth]{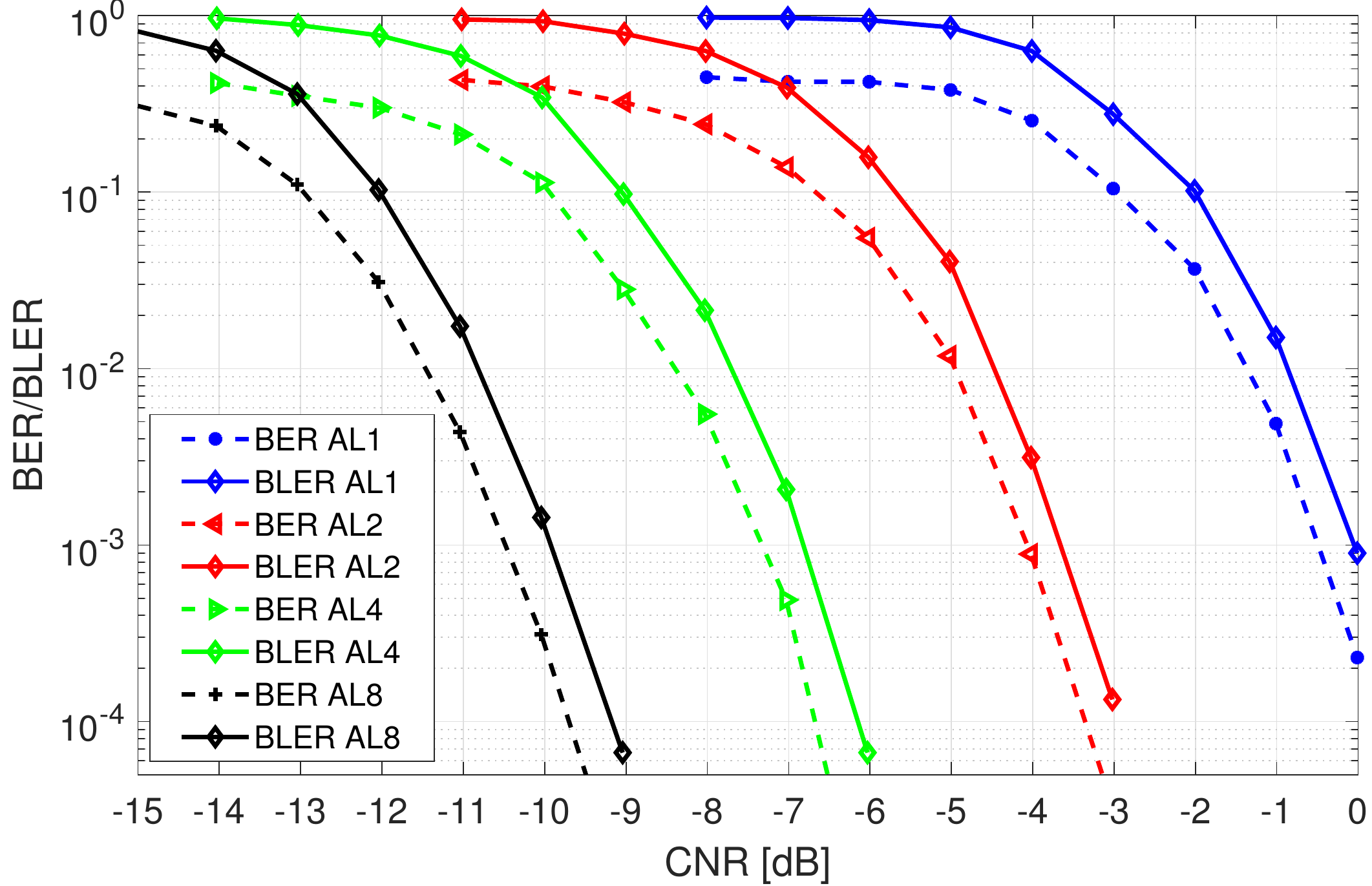}
	\caption{BER/BLER vs. CNR (dB) for TDL-C, ideal channel estimation.}
	\label{CDLC}
\end{figure} 

Results for AWGN are shown in Fig. \ref{AWGNall}. From the figure, we conclude that a higher aggregation level generally gives more protection level to the codewords, which is reflected on the required CNR for both LTE-eMBMS and NR-PTP solutions. at the expense of more occupied bandwidth. Due to the different CRC and aggregation level for LTE and NR, it is not easy to show a fair comparison, although generally polar codes outperform tail-biting encoding. A suitable Quality of Service (QoS) metric can be a BLER lower than 0.1\% for a reliable broadcasting transmission \cite{D31}. Comparing under the same aggregation level 1, NR requires about 2.8 dB less than LTE to achieve this criterion.

\subsubsection{TDL Channel Models Considered in IMT-2020 Scenarios}
In this subsection, the BLER performance for the IMT-2020 scenarios is presented with perfect channel estimation. Fig. \ref{CDLA} shows the results for the TDL-A channel model with 30 ns of delay spread and 3 km/h user speed. Fig. \ref{CDLC} depicts the results for the TDL-C channel model, for a rural scenario with 300 ns of delay spread and 30 km/h user speed. 
From the results, we can see that under perfect channel estimation, a higher movement speed is equivalent to a larger Doppler diversity. This involves that the CNR requirement for each aggregation level in the TDL-C channel (30 km/h speed) significantly outperforms the requirement with TDL-A channel (3 km/h speed). Compared to AWGN, with low aggregation levels equivalent to high code rates, the BLER performance for both TDL-A and TDL-C channels are worse. However, because of the fixed codeword length, with higher aggregation levels the code rate dramatically decreases and the TDL channel performance is almost aligned with AWGN in this case.

\subsection{Real Channel Estimation}
In this work, we only study the real channel estimation evaluation with 5G New Radio. We assume a RB-based 2-dimensional linear channel estimation, and we follow the frame structure introduced in Sec.\ref{sec:FS}. Regarding the DMRS signals transmitted in each RB for channel estimation, we have:

	\begin{itemize}
		\item Frequency domain: DMRS allocated every 4 subcarriers.
		\item Time domain: the number of DMRS symbol depends on the $N^{CORESET}_{symb}$ value, which is determined by PCFICH and can be 1, 2 or 3.
	\end{itemize}

In order to estimate the channel, the two-dimensional (frequency and time) sampling must satisfy:
	\begin{itemize}
		\item Frequency domain: the sampling rate must be faster than or equal to the channel's maximum delay spread.
		\item Time domain: the sampling rate must be greater than or equal to the channel's maximum Doppler spread.
	\end{itemize}

The maximum distance between two time domain DMRS symbols is given by:

\begin{equation}
n \leq \frac{1}{2*T_s*d_{max}},
\end{equation}

\noindent where $T_s$ and $d_{max}$ represent the symbol duration and maximum Doppler spread respectively. Due to the fact that DMRS signals cover the whole time domain, considered pilots are sufficient to capture the time-variation of the channel with potentially any user speed. On the other hand, the frequency domain channel estimation depends on the maximum channel delay spread, and the maximum distance between two frequency domain pilots is given by:

\begin{equation}
m \leq \frac{T_s}{2*\tau _{max}},
\end{equation}

where $\tau_{max}$ represents the maximum channel delay spread. In this work we assume 15 kHz subcarrier spacing and $m=4$ as shown in Fig. \ref{Fig_Framestructure}, which gives: $ T_s =1/\Delta_f = 66.7 \mu s$. Therefore, the maximum channel delay spread that can be tolerated is $\tau_{max}\leq T_s/2m \approx 8.33\mu s$, greater than the one from the selected test channels TDL-A and TDL-C. 

\begin{figure}[t]
	\centering
	\includegraphics[width=0.49\textwidth]{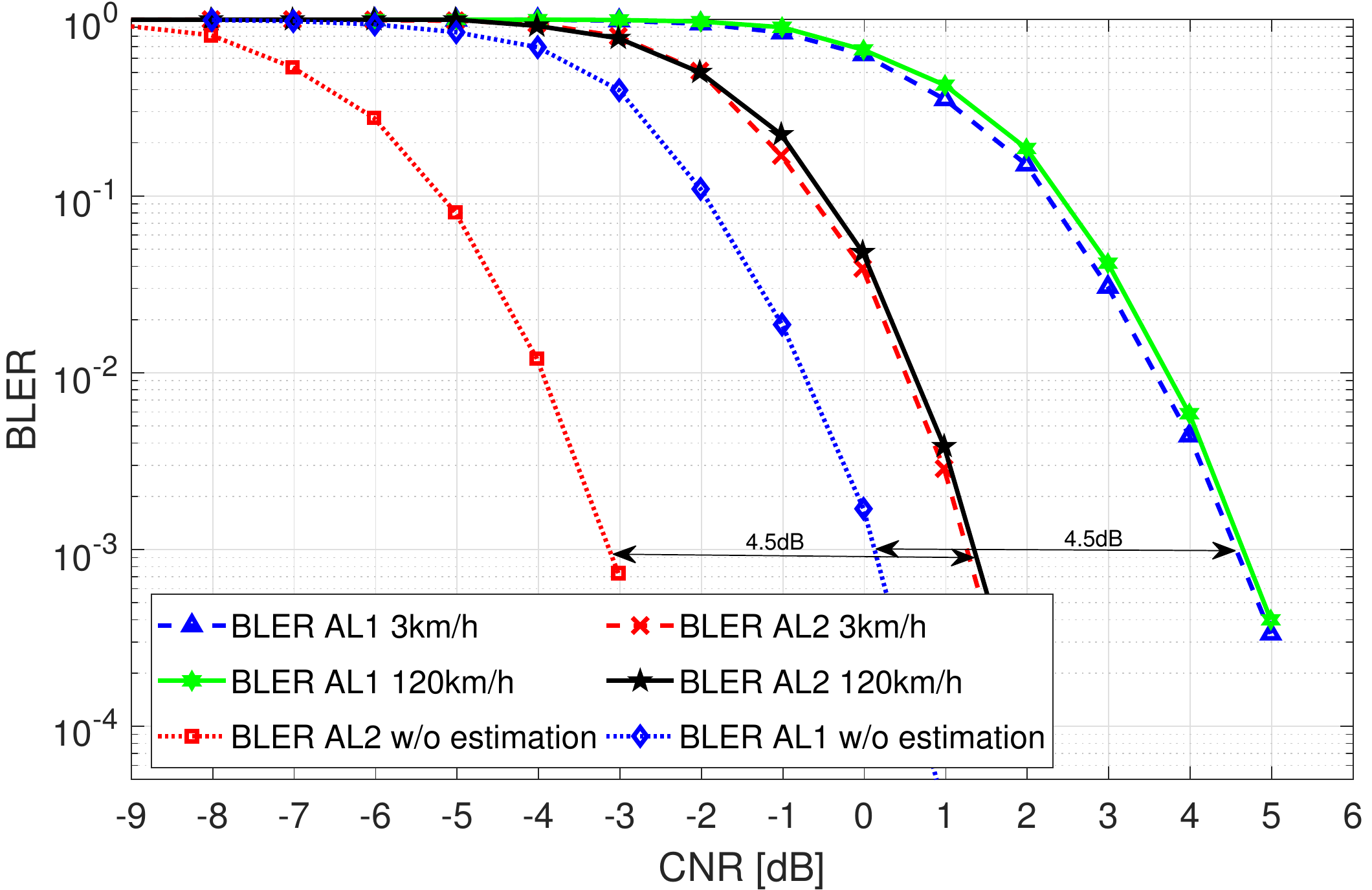}
	\caption{BICM BER/BLER vs. CNR (dB) for TDL-A channel with ideal and real channel estimation.}
	\label{Fig_chest}
\end{figure} 

Fig. \ref{Fig_chest} shows simulation results with real and ideal channel estimation, for TDL-A channel. The difference in terms of BLER performance vs required CNR  is about 4.5 dB for both aggregation levels 1 and 2. As expected, this is due to the noise effect during the channel estimation. Note that the legend in Fig. \ref{Fig_chest} includes the term 'w/o estimation' to indicate the simulations obtained without channel estimation, and the rest have used real channel estimation. Moreover, we can see that different user speeds of 3 km/h and 120 km/h almost have no impact in the BLER performance, which reflects the rationality of the pilot distribution.

\section{Conclusion}\label{sec:conclusion}
In this paper, the DCI generation for both LTE eMBMS and NR PTP technologies has been explained. A detailed technical overview of the transmission of control information in both systems has been also covered. The PDCCH performance has been analysed for AWGN, as well as for TDL-A and TDL-C channel models. The discussions and simulation results obtained in this paper can be used as a benchmark to evaluate the end-to-end system performance of NR PTP, and more importantly, to propose suitable control channel configurations for possible solutions towards NR PTM transmissions. A potential way forward includes first the definition of a new DCI format for NR PTM. Secondly, most of the PDSCH performance evaluation assumes perfect PDCCH signal recovery, so one can also extend this work onto the performance evaluation of the data channel in the presence of imperfect PDCCH transmissions.






%
\bibliographystyle{IEEEtran}
\bibliography{reference}

\end{document}